\documentclass[conference]{IEEEtran}
\IEEEoverridecommandlockouts
\usepackage{cite}
\usepackage{amsmath,amssymb,amsfonts}
\usepackage{algorithmic}
\usepackage{graphicx}
\usepackage{textcomp}
\usepackage{xcolor}
\usepackage{tikz}
\usepackage{comment}
\usepackage{stfloats}  
\def\BibTeX{{\rm B\kern-.05em{\sc i\kern-.025em b}\kern-.08em
    T\kern-.1667em\lower.7ex\hbox{E}\kern-.125emX}}
\usepackage{enumitem,amssymb}
\usepackage{placeins}

\usepackage{threeparttable}
\usepackage{booktabs}
\usepackage{array}

\usepackage{subcaption}
\usepackage[table]{xcolor}

\newlist{todolist}{itemize}{2}
\setlist[todolist]{label=$\square$}

\newcommand{\review}[1]{{\textcolor{black}{#1}}}
\newcommand{\cameraready}[1]{{\textcolor{black}{#1}}}

\begin{document}
\title{A Parallel Ultra-Low Power Silent Speech Interface based on a Wearable, Fully-dry EMG Neckband\\
\thanks{The authors acknowledge support from the ETH-Domain Joint Initiative program (project UrbanTwin).}
}
\author{
\IEEEauthorblockN{
    Fiona Meier\IEEEauthorrefmark{1}, 
    Giusy Spacone\IEEEauthorrefmark{1}, 
    Sebastian Frey\IEEEauthorrefmark{1},  
    Luca Benini\IEEEauthorrefmark{1}\IEEEauthorrefmark{2}, 
    Andrea Cossettini\IEEEauthorrefmark{1}
}
\IEEEauthorblockA{\IEEEauthorrefmark{1}
Integrated Systems Laboratory, ETH Zurich, Z{\"u}rich, Switzerland, \IEEEauthorrefmark{2}DEI, University of Bologna, Bologna, Italy}
}

\makeatletter
\def\ps@IEEEtitlepagestyle{%
  \def\@oddhead{}
  \def\@evenhead{}%
  \def\@oddfoot{%
    \vbox to0pt{\vss
      \hbox to\textwidth{%
        \parbox[t]{\textwidth}{\centering\scriptsize
          \copyright 2025 IEEE.  Personal use of this material is permitted.  Permission from IEEE must be obtained for all other uses, in any current or future media, including reprinting/republishing this material for advertising or promotional purposes, creating new collective works, for resale or redistribution to servers or lists, or reuse of any copyrighted component of this work in other works.
        }%
      }%
    }%
  }%
  \def\@evenfoot{}%
}
\makeatother

\maketitle
\begin{abstract}
We present a wearable, fully-dry, and ultra-low power EMG system for silent speech recognition, integrated into a textile neckband to enable comfortable, non-intrusive use. The system features 14 fully-differential EMG channels and is based on the \cameraready{BioGAP-Ultra} platform for \review{ultra-low power (22 mW)} biosignal acquisition and wireless transmission. We evaluate its performance on eight speech commands under both vocalized and silent articulation, achieving average classification accuracies of 87$\pm$3\% and 68$\pm$3\%respectively, with a 5-fold CV approach. To mimic everyday-life conditions, we introduce session-to-session variability by repositioning the neckband between sessions, achieving leave-one-session-out accuracies of 64$\pm$18\% and 54$\pm$\review{7}\%  for the vocalized and silent experiments, respectively. These results highlight the robustness of the proposed approach and the \review{promise} of energy-efficient silent-speech decoding.

\end{abstract}


\vspace{-0.3cm}
\section{Introduction}

Silent speech interfaces (SSIs) are assistive technologies that enable human-machine interaction through speech-decoding without vocalization \cite{freitas_introduction_2017, gonzalez2020silent}. These systems are particularly valuable for individuals with speech impairments \cite{gonzalez2020silent}, as well as in situations requiring silent communication, such as in noisy environments, or to guarantee enhanced privacy \cite{denby2010silent, krishna2019speech}.

SSIs can be broadly categorized by the sensing modality \cite{gonzalez2020silent}: brain activity, muscular activity, speech, or articulatory activity. 
In this context, surface electromyography (EMG) appears as the preferred approach for muscle-based SSIs, due to its non-invasive nature \cite{gonzalez2020silent} and thanks to the early onset of EMG signals (60 ms before
articulatory motion) \cite{netsell1974neural}.

While most prior works are based on wet electrodes attached to the face of the subjects \cite{diener2015direct}, this approach poses challenges for social acceptability and long-term use. \review{Thus,} recent efforts are focused on fully-dry EMG solutions.

One notable example of EMG-based non-invasive SSI is AlterEgo\cite{kapur_alterego_2018}, which demonstrated an accuracy as high as 92.01\% for a digit vocabulary. However, the setup is based on EMG measurements on the face, which hinder its widespread adoption due to stigmatization. While epidermal devices appear as a promising solution for concealed face-based measurements \cite{liu2020epidermal}, their integration with compact and ultra-low-power readout electronics is still not demonstrated.

More recent systems have explored alternative form factors for concealed EMG acquisition \cite{tang2025wireless, kang2025wearable, wu_towards_2024}.
In \cite{tang2025wireless}, a 4-channel EMG interface is integrated into headphones, achieving 96\% accuracy on 10 command words. However, it \review{relies} on silver paste for electrode adhesion, unsuitable for mass deployment, and power efficiency was also not fully characterized.
Kang et al. \cite{kang2025wearable} proposed a neck-worn EMG-piezoelectric system, demonstrating improved classification accuracy thanks to sensor fusion \review{compared to using a single modality}. However, the use of wet electrodes limit \review{application on everyday life usage scenarios}.
Wu et al. \cite{wu_towards_2024} introduced a wireless and stretchable EMG neckband with 10 differential EMG channels and a low-power wireless acquisition platform, achieving an accuracy of 92.7\% across 11 words. However, the system uses a 15-bit ADC (lower than the 24-bit commonly used for EMG), \review{is not fully-dry (requires a wet reference electrode),} and lacks substantial on-board processing capabilities. Our work takes inspiration from \cite{wu_towards_2024}, aiming to improve on these limitations.

Beyond hardware constraints, most existing studies do not account for every-day usage conditions such as device removal and repositioning between sessions, which can significantly affect signal quality and repeatability.

We address these gaps with the following contributions:
\begin{itemize}
    \item Design of a fully-dry, textile-based neckband with 14 channels along the neck, enabling comfortable and non-stigmatizing EMG acquisition. The system integrates the \cameraready{BioGAP-Ultra\cite{frey_biogap2_2025}} platform for data acquisition with 24-bit resolution and continuous BLE streaming at ultra-low-power ($22.2\,\text{mW}$).
    \item Demonstration of accurate 8-word classification for both vocalized (87$\pm$3\%) and silent (68$\pm$3\%) articulation using a Random Forest classifier with a 5-fold CV approach.
    \item Demonstration of robustness to real-world variability by removing and repositioning the setup before each session. Leave-one-session-out (LOSO) analyses show robust performance (64$\pm$18\% vocalized, \review{54$\pm$7\%}  silent), well above chance (12.5\% for 8 classes).
\end{itemize}


\section{Materials and Methods}

\subsection{EMG neckband design}
\begin{figure}
    \centering
\includegraphics[width=1\linewidth]{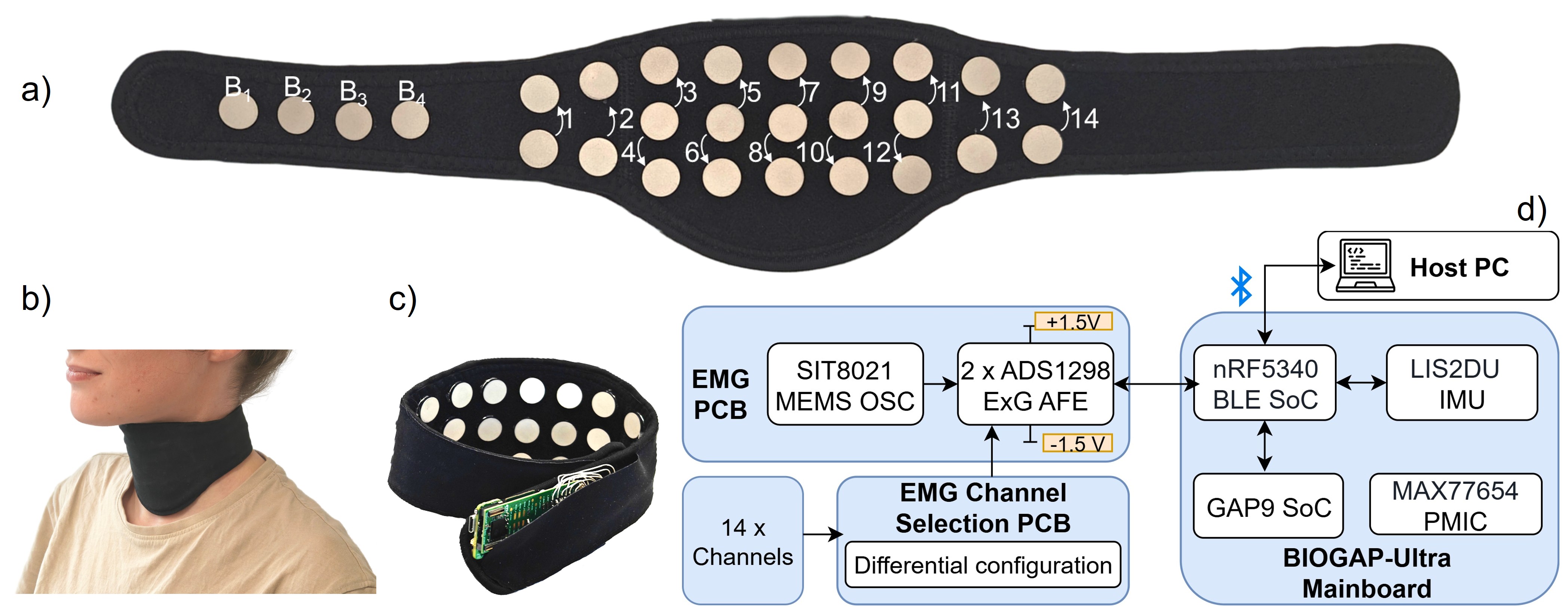}

    \caption{\textbf{a)} Neckband with highlighted distribution of the 14 differential channels  ($B_{1..4}$ are the bias channels). \textbf{b)} Neckband worn by a healthy volunteer. \textbf{c)} Neckband and acquisition electronics. \cameraready{d) Block diagram of the acquisition system.}}
    \label{fig:setup}
    \vspace{-0.7cm}
\end{figure}

The proposed system is based on a modified commercial textile neckband \cite{neckband}, which offers adjustable sizing, durable fabric, and secure Velcro fastening. As shown in Fig.~\ref{fig:setup}, 27 snap fasteners were sewn onto the band to allow easy connection of fully-dry, reusable biopotential passive electrodes (SoftPulse, Datwyler \cite{datwyler_softpulse}). Electrodes are arranged in multiple rows, with a higher density at the front of the neck to target muscles more involved in speech production \cite{hoerter_anatomy}. A fully-differential configuration is implemented to suppress common-mode noise \cite{merletti2020tutorial}. The central row is shared between the top and bottom rows to implement differential channels, resulting in a total of ten differential pairs. Two additional differential channels are placed laterally on each side of the neck. Ground is provided via four electrically shorted electrodes located at the back of the neck. 

\subsection{Data acquisition platform}
\vspace{-0.1cm}
The neckband is interfaced with \cameraready{BioGAP-Ultra \cite{frey_biogap2_2025}}, a parallel ultra-low-power biosignal acquisition platform integrated in the rear portion of the textile band. \cameraready{BioGAP-Ultra} consists of a baseboard incorporating GAP9, a state-of-the-art ultra-low-power System on Chip (SoC) for edge-AI, and a \review{Bluetooth-low-energy (BLE)-capable} microcontroller for wireless raw data streaming \review{(NRF5340, Nordic Semiconductor)}. The baseboard connects to a biopotential acquisition board equipped with two ADS1298 \review{(Texas Instruments)} analog front ends (AFE), enabling simultaneous acquisition from 16 differential EMG channels. \cameraready{The complete system's description is provided in \cite{frey_biogap2_2025}}.

All signals were acquired with an amplification gain of 12 and a sampling rate of 500 samples per second (SPS).

\subsection{Data collection protocol}
\vspace{-0.1cm}
Data were collected from a single healthy volunteer performing a speech task composed of eight target words: \textit{up}, \textit{down}, \textit{left}, \textit{right}, \textit{forward}, \textit{backward}, \textit{go}, and \textit{stop}. These words were selected for their relevance in potential control applications (e.g., robotic navigation). The protocol included two conditions: vocalized speech and silent articulation.

Before each session, the skin of the participant was cleaned with water. 
Word prompts were presented using a Python script running on the same laptop that received the EMG data from BioGAP, ensuring synchronized labeling. For each prompt, the participant had four seconds to articulate the word, followed by a one-second rest. The word order was randomized to prevent the possibility of anticipation.

Data were acquired over three separate sessions yielding 3360 utterances for each experiment type (vocalized or silent). Each session consisted of 7 batches, with each batch containing 20 repetitions of the eight words (160 utterances per batch, 1120 per session). Each batch lasted approximately 14 minutes, with a 5 minutes break between batches. 

To simulate realistic use, the neckband was removed and repositioned before each session.
\subsection{Data pre-processing and feature extraction}
\vspace{-0.1cm}
Raw EMG data are filtered with a $4^{th}$-order zero-phase high-pass filter with a $20 \text{Hz}$ cutoff frequency, followed by a $50\text{Hz}$ notch filter to eliminate powerline interference. 

\begin{figure}
    \centering\includegraphics[width=0.49\textwidth]{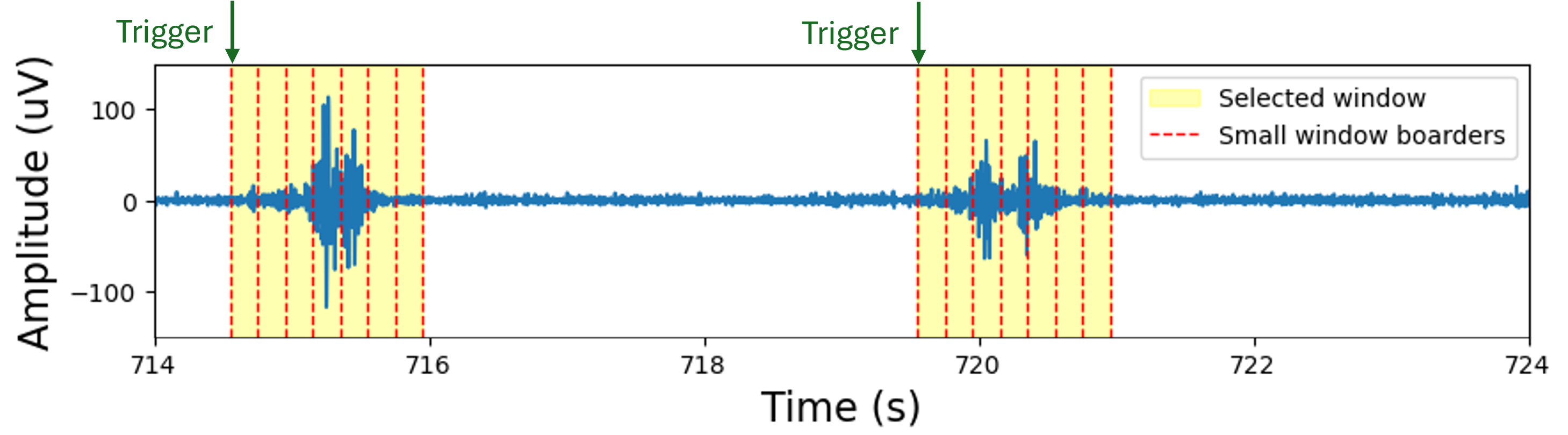}

    \caption{\cameraready{EMG signal during the vocalized experiment. The green arrows correspond to the prompt from the GUI, and the yellow region shows the window considered for classification, divided into segments for feature extraction.}}
    \vspace{-0.6cm}
    \label{fig:windowing}
\end{figure}

Out of the four seconds of data measured for each utterance, we restrict the data analysis to only the first $1.4\text{s}$, since each articulation typically takes approximately one second (see Fig.~\ref{fig:windowing}). These $1.4\text{s}$ of data are then segmented into seven $200\text{ms}$ non-overlapping windows. For each window, we extract \review{the same} time-domain and frequency-domain features \review{as in} \cite{liu2020epidermal, wu_towards_2024, eddy_libemg}, \review{given their demonstrated suitability for this classification task}.
Time domain features include root mean square (RMS), max, min, standard deviation, variance, mean, 25th percentile, 75th percentile, zero-crossing rate) and the mean and standard deviation of third-level Daubechies (db4) wavelet coefficients. Frequency-domain features include  mean and peak frequency, total and mean power, second- and third-order spectral components, and frequency distribution. Features from all seven windows spanning the $1.4\text{s}$ articulation interval are concatenated to represent each utterance. \review{Feature importance analysis revealed no dominant feature, suggesting to keep the full set.}
\vspace{-0.3cm}
\subsection{Training strategy and evaluation procedure}
\vspace{-0.1cm}
A Random Forest classifier is used for the 8-class word classification (\textit{scikit-learn} \cite{scikit-learn} with default parameters). \review{We chose this model for its good performance with minimal parameter tuning and its interpretable feature importance.} We consider the following evaluation strategies:

\textbf{Session-specific model.} For each session, a 7-fold cross-validation (CV) is performed across its seven batches. Each fold uses six batches for training and one for testing. Results are reported as the mean and standard deviation of the accuracy per label, as well as the overall accuracy metric.

\textbf{Global model (5-fold CV)} We aggregate the data of all sessions and randomly shuffle them. We then perform a stratified 5-fold CV. This approach ensures that each fold sees data from all acquisition sessions, hence accounting for the repositioning of the sensors, while maintaining balance between the classes. We report per-label and overall accuracy (mean and standard deviation) across the 5 folds.

\textbf{Leave-one-session-out (LOSO)}. The model is trained on two sessions and tested on the remaining third session. This analysis evaluates the model’s robustness to electrode repositioning. We report per-label and overall accuracy (mean and standard deviation).
\section{Results and Discussion}
\vspace{-0.1cm}
\begin{figure}[t] 
    \centering
    \begin{subfigure}{0.95\linewidth}

        \centering        \includegraphics[width=1\linewidth]{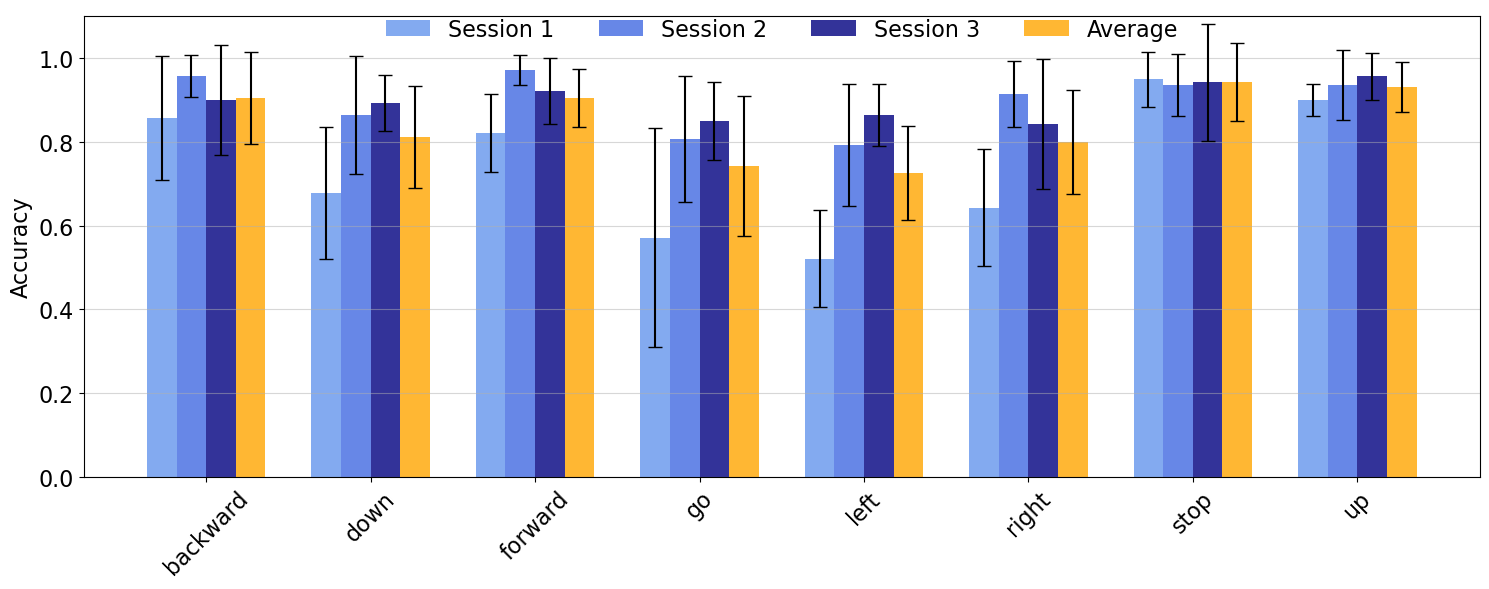}
        \vspace{-0.7cm}
        \caption{Session-specific model}
        \label{fig:subfig1}
    \end{subfigure}

    \begin{subfigure}{0.92\linewidth}
        \centering
        \includegraphics[height=3.7cm]{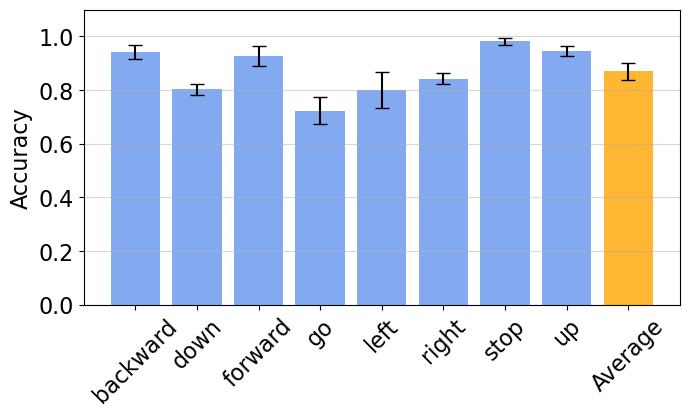}
        \vspace{-0.4cm}
        \caption{Global Model (5-fold CV)}
        \label{fig:subfig2}
    \end{subfigure}

    \begin{subfigure}{0.92\linewidth}
        \centering
        \includegraphics[height=3.7cm]{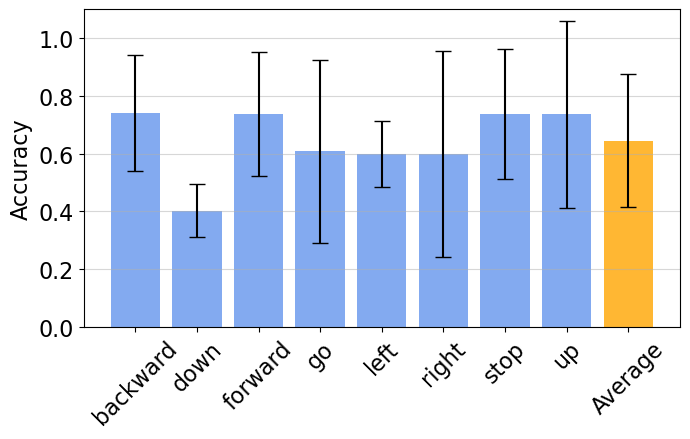}
        \vspace{-0.5cm}
        \caption{LOSO}
        \label{fig:subfig3}
    \end{subfigure}

    \caption{Accuracy results for the vocalized experiments.}
\vspace{-0.5cm}\label{fig:vocalized_accuracy}
\end{figure}
\begin{figure}[h] 
    \centering
    \begin{subfigure}{0.95\linewidth}
        \centering

        \includegraphics[width=1\linewidth]{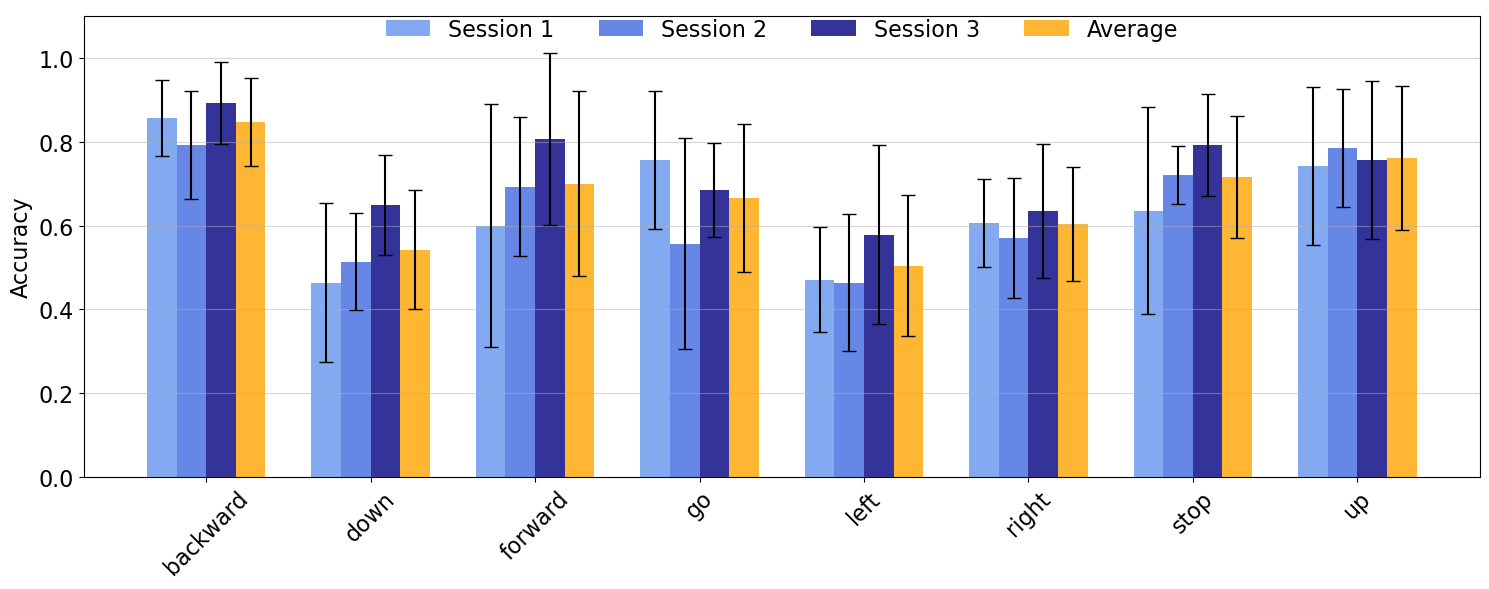}
        \vspace{-0.7cm}
        \caption{Session-specific model}
        \label{fig:subfig1}
    \end{subfigure}

    \begin{subfigure}{0.92\linewidth}
        \centering
        \includegraphics[height=3.7cm]{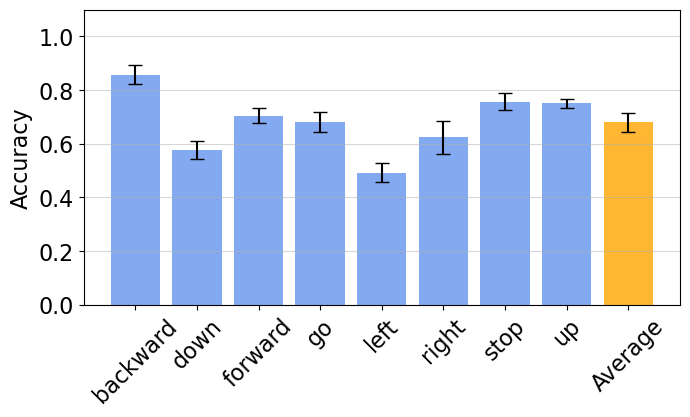}
        \vspace{-0.4cm}
        \caption{Global Model (5-fold CV)}
        \label{fig:subfig2}
    \end{subfigure}

    \begin{subfigure}{0.92\linewidth}
        \centering
        \includegraphics[height=3.7cm]{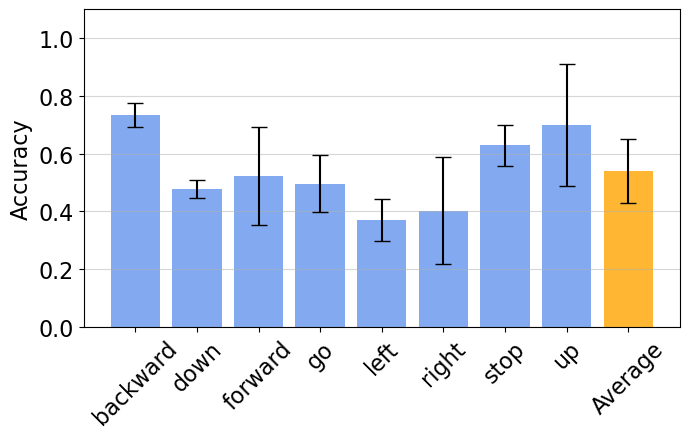}
        \vspace{-0.6cm}
        \caption{LOSO}
        \label{fig:subfig3}
    \end{subfigure}
    \caption{Accuracy results for the silent experiments.}
    \label{fig:silent_accuracy}
    \vspace{-0.6cm}
\end{figure}

\subsection{Vocalized speech experiments}
\vspace{-0.1cm}
Fig.~\ref{fig:vocalized_accuracy} (top) shows the per-session accuracy for the vocalized experiments. On average, an accuracy of \review{85$\pm$7\%} is achieved for the 8-class classification problem, with a lower accuracy in the first session (74\%), which we attribute to the inexperience of the subject during this first-time experiment. Fig.~\ref{fig:vocalized_accuracy} (middle) shows the results for the global model, which achieves a total accuracy of 87$\pm$3\%.
Fig.~\ref{fig:vocalized_accuracy} (bottom) shows the LOSO results \cameraready{(accuracy 64$\pm$18\%); the drop compared to the global model highlights the challenges introduced by repositioning the device.}
Nevertheless, \cameraready{results} stand well above the random guess (threshold: 12.5\%).
\vspace{-0.2cm}
\subsection{Silent articulation experiments}
\vspace{-0.2cm}
Fig.~\ref{fig:silent_accuracy} shows the per-session (top), global (middle), and LOSO (bottom) results for the silent articulation experiments. We observe a significant drop in the accuracy compared to the vocalized experiments (68$\pm$3\% vs 87$\pm$3\% \review{for the Global Models}), which we attribute to the reduced number of muscles involved in the articulation when the subject does not vocalize the sound. The results are still much higher than a random guessing.
For the most challenging scenario, i.e. silent articulation in a leave-one-session-out setting, we observe an average accuracy of \review{54$\pm$7}\%.

\vspace{-0.1cm}
\section{Conclusion}
\vspace{-0.1cm}
\cameraready{We present a fully-dry, wearable interface for silent speech decoding based on a textile EMG neckband. Powered by BioGAP-Ultra, our system can sustain continuous data acquisition and wireless transmission via BLE with a power consumption of only 22.2 mW.}

We validate the system on both vocalized and silent articulation of eight target words, achieving an average cross-validated classification accuracy for 8 classes of 87$\pm$3\% and 68$\pm$3\%, respectively. \review{The vocalized accuracy is only slightly lower than the 93\% of \cite{wu_towards_2024}, which, however, requires a wet reference (and uses a different vocabulary).} \review{Our results are also comparable to the 89.6\% accuracy of  \cite{liu2020epidermal}, obtained for a similar HMI vocabulary (with a lower number of classes and with a non-wearable system).} To assess usability for everyday-life conditions, we introduced session-to-session variability by repositioning the neckband between sessions. In this scenario, leave-one-session-out analyses yield average accuracies of 64$\pm$18\% and \review{54$\pm$7}\%  for vocalized and silent articulations, respectively.

\cameraready{Future work will focus on increasing channel count with miniaturized electrodes, integrating additional sensing modalities to enhance robustness, and extending both the vocabulary set and cohort size for further validation.}

\cameraready{In conclusion, these} results demonstrate the feasibility of \cameraready{wearable}, fully dry EMG-based SSIs at ultra-low power consumption.

\vspace{-0.15cm}
\section*{Acknowledgment}
\vspace{-0.2cm}We thank A. Blanco Fontao, H. Gisler, T. Quanbrough (ETH Z{\"u}rich) for technical support. 

\bibliographystyle{IEEEtran}
\bibliography{bibliography}
\end{document}